\documentclass[12pt,epsf,epsfig,psfig]{article}
\usepackage{graphicx}
\usepackage{epsfig}
\usepackage{cite}
\def\e{\epsilon}

\def\be{\begin{equation}}
\def\ee{\end{equation}}

\def\lsim{\raise0.3ex\hbox{$<$\kern-0.75em\raise-1.1ex\hbox{$\sim$}}}
\def\gsim{\raise0.3ex\hbox{$>$\kern-0.75em\raise-1.1ex\hbox{$\sim$}}}

\def\NP{{ Nucl.\ Phys.\ }}
\def\PL{{ Phys.\ Lett.\ }}
\def\PR{{ Phys.\ Rev.\ }}

\def\PRL{{ Phys.\ Rev.\ Lett.\ }}

\def\ZP{{ Z.\ Phys.\ }}
\def\EP{{ Eur.\ Phys.\ J.\ C}}
\def\J{{$J/\psi$}}

\begin{document}

June 2012 \hfill BI-TP 2012/23

\vskip 2cm

\centerline{\Large \bf Quark Matter and Nuclear Collisions$^{\large *}$}

\vskip0.5cm

\centerline{\large \bf A Brief History of} 

\medskip

\centerline{\large \bf Strong Interaction Thermodynamics} 

\vskip 1cm 

\centerline{\large \bf Helmut Satz} 

\bigskip

\centerline{Fakult\"at fur Physik, Universit\"at Bielefeld}

\medskip

\centerline{Postfach 100 131, D-33501 Bielefeld, Germany}

\vskip3cm

\centerline{\bf \large Abstract:}

\bigskip

The past fifty years have seen the emergence of a new field of research
in physics, the study of matter at extreme temperatures and densities.
The theory of strong interactions, quantum chromodynamics (QCD), predicts 
that in this limit, matter will become a plasma of deconfined quarks and
gluons -- the medium which made up the early universe in the 
first 10 microseconds after the big bang. High energy nuclear collisions 
are expected to produce short-lived bubbles of such a medium in the 
laboratory. I survey the merger of statistical QCD and nuclear collision 
studies for the analysis of strongly interacting matter in theory 
and experiment.  

\vfill

\hrule

\medskip

${\large *}$ Opening Talk at the {\sl 5th Berkeley School on 
Collective Dynamics in High Energy Collisions}, LBNL Berkeley/California,
May 14 - 18, 2012.

\newpage

~~~\vskip2cm

\centerline{\bf \Large Contents}

\leftskip1.5cm

\vskip1cm

1.\ Denser than Nuclei: 
Neutron Stars, Big Bang, Heavy Ion 
Collisions

\bigskip

2.\ All those Resonances: Hagedorn's Legacy

\bigskip

3.\ The Conjecture of Lucretius: Quark Confinement

\bigskip

4.\ A Shift of Paradigm: 
The Computer Simulation of Lattice QCD 

\bigskip

5.\ The Little Bang: 
Making Matter in Collision

\bigskip

6.\ The Abundance of the Species:
Universal Hadrosynthesis

\bigskip

7.\ Melting of Quarkonia: The QGP Temperature

\bigskip

8.\ Quenching of Jets: The QGP Density

\bigskip

9.\ Horizons: Limits of Communication

\leftskip0cm

\newpage

\vskip2cm

\section{\large Denser than Nuclei:\\
\hfill Neutron Stars, Big Bang, Heavy Ion 
Collisions}

\vskip0.5cm

How dense can matter be - is there a limit? This question has been around
for quite a while. A celebrated answer was triggered by Sir Walter Raleigh,
who was wondering how one could best stack the cannon balls on his ships. He
asked his expert on board, the English mathematician Thomas Harriot, who
passed the question on to his German colleague Johannes Kepler. Kepler,
in 1611, suggested that the ordered stack of hard balls, still used today by 
grocers to display oranges, is the greatest possible density for such objects;
it filled the volume to a fraction of $\pi/\sqrt 18\simeq 0.74$. In 1831,
Carl Friedrich Gauss proved that Kepler's conjecture is in fact correct, 
for orderly packing. But if we simply pour balls into a 
container, we never reach such a density, since the randomly falling 
balls do not arrange themselves in such an orderly pattern. The maximum 
density of the disordered medium is much more difficult to calculate -- in 
fact, that this density is always less than that of the Kepler pyramid was 
established only recently by the American mathematician Thomas C.\ Hales, 
using a computer-aided proof. And when the crate is filled, we can still 
increase the density a little by shaking the container for a while. Such
random close packing leads to a density limit of 0.65 -- about 10 \% less 
dense than the orderly stacked pile.
 
\begin{figure}[htb]
\centerline{\epsfig{file=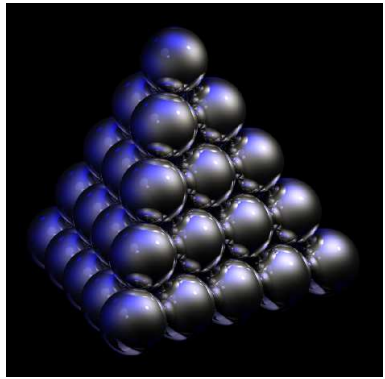,width=5cm}}
\caption{Stacking cannon balls}
\label{cannon}
\end{figure}

\medskip

The nuclei of heavy atoms are the densest matter found on Earth, and they
are also the only strongly interacting matter here. Standard nuclear density 
is 0.16/fm$^3$; with a nucleon radius of about 0.8 fm, that means about 35 \% 
of the nuclear volume is filled with nucleons, 65 \% remain empty. So the 
nucleons can still rattle around a little inside the nucleus. But they are 
``jammed'', confined 
to remain in a restricted local environment. And twice normal nuclear density 
comes already close to Kepler's conjecture for the orderly packing limit, 

\medskip

Neutron stars are the result of stellar collapse under gravity, occurring
when the nuclear fuel has burnt up and there is no further thermal pressure 
to balance the gravitational force. If the mass of the star is large
enough, gravity will force electrons and protons to undergo inverse beta-decay
($p + e^- \to n$)
and fuse to neutrons, so that now a combination of nucleon repulsion and
Fermi pressure stops gravity from causing further compression.
This, however, means that the mass can also not be too large: otherwise 
gravity will win after all and turn the dead star into a black hole. 
The density of a neutron star increases in going from the crust to the 
core, and present estimates of core densities go up to five times normal
nuclear density. That is well above all close packing limits, random as
well as orderly, and so it must mean that the nucleons suffer considerable
squeezing and/or overlap. Which raises the question of whether this allows 
them to remain as nucleons or whether the dense neutron star cores consists 
of deconfined quarks.

\medskip
  
The expansion of the present universe, as attested by Hubble's law as well
as by the cosmic background radiation, allows us to extrapolate its density
back to early times, close to its origin in a hot Big Bang.
The relation between the age of the universe and its energy density is
is given by
\be
t = {1 \over H(t)} = \sqrt{{3 \over 8\pi G \epsilon(t)}},
\label{ex-time}
\ee
where $H(t)$ is the Hubble parameter, $G$ is the gravitation constant and 
$\e(t)$ the cosmic energy density. Considering the instant of the big bang 
as starting point, we can determine the time needed to reach, on a cosmic 
level, the energy density $\e \simeq 0.5$ GeV/fm$^3$, found inside a single 
nucleon. With the value $G=6.708 \times 10^{-39}~{\rm GeV}^{-2}$, it becomes
\be
t_q \simeq 10^{-5}~{\rm s},
\ee
so that in its first ten microseconds, our universe was in a state for which
the existence of individual hadrons is not really conceivable. Today we
take that primordial medium to consist of deconfined quarks and define the 
time up to $t_q$ as the {\sl quark era}. Only at the end of this era 
hadrons were formed, and there appeared for the first time the stage 
for today's complex cosmic structure, the physical vacuum.

\medskip

\begin{figure}[htb]
\vspace*{-0.5cm}
\centerline{\epsfig{file=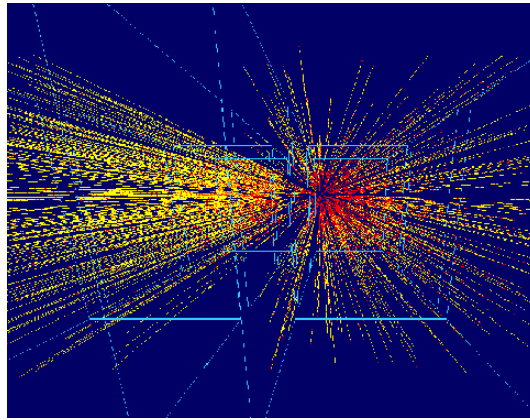,width=5.5cm,angle=-90}}
\caption{Particle production in the collision of two gold nuclei at 
$\sqrt s =  200$ GeV, from the STAR experiment at RHIC,
Brookhaven National Laboratory}
\label{RHIC}
\end{figure}

Given the density of a single heavy nucleus: if we collide two of them at
high energy, they will either compress each other or overlap, and the 
resulting medium must be, at least for a short time, of considerably higher 
nucleon density, and hence also of rather high energy density.
And indeed, the collision of two gold nuclei at a center of mass
energy of 200 GeV turns this high energy density into the production of 
hundreds of secondary hadrons, see Fig.\ \ref{RHIC}. 

Tracing all these hadrons back to the interaction region of the two colliding
nuclei, it is clear the the energy density there must have been very much
higher than the 0.5 GeV/fm$^3$ inside a single nucleon. Whether it is 
meaningful to call this dense medium ``matter'' is another issue, subject 
of many studies, both theoretical and experimental.

\section{\large All those Resonances: Hagedorn's Legacy}

\vskip0.3cm

In the middle of the last century, the ultimate constituents of matter seemed
fairly clear: protons and neutrons giving the mass of nuclei, the almost
massless electrons to assure electric neutrality. The short-range {\sl strong}
force, required to bind the nucleons, led Yukawa to introduce the pion
as the boson mediating the interaction, and challenged experimentalists
to look for it in cosmic rays as well to produce it in the laboratory.
Proton-proton collisions indeed did that, but also much more: they opened
a Pandora's box containing, by now, an almost unlimited number of new strongly 
interacting elementary or not so elementary particles, {\sl hadrons} -- with 
more than a thousand entries in the Particle Data List, according to the 
latest count. New discrete, non-kinetic quantum numbers were needed: isospin,
strangeness, charm and beauty, and for a given such quantum number, ever 
growing towers of resonances of different angular momenta and parities
were observed.

\medskip

This multitude of hadrons had two consequences which changed the course of 
physics. One
was in its initial concept not really revolutionary, though it eventually
provided us with the theory of strong interactions, {\sl quantum 
chromodynamics}. The idea was simply that so many elementary particles
must be compound states of fewer and simpler constituents, and such
{\sl reductionism} has been with us ever since the atoms of ancient 
Greece. We shall return to the results for strong interaction physics 
in the next section.

\medskip

The second outcome was indeed totally new, and the basic idea came from
Rolf Hagedorn \cite{Hage1,Hage2}. He first asked whether the composition of 
resonances might not
follow a self-similar pattern, so that a resonance of a given mass would
consist of lighter resonances constructed according to the same composition
law. In Hagedorn's words, ``fireballs consist of fireballs which consist o
fireballs, and so on...''. A more poetic summary was already given around
the middle of the nineteenth century by the English mathematician Augustus 
de Morgan (inspired by some lines of Jonathan Swift), 

\begin{center}
{\sl Great fleas have little fleas upon their backs to bite 'em.

And little fleas have lesser fleas, and so ad infinitum.

And the great fleas themselves, in turn, have greater fleas to go on,

While these again have greater still, and greater still, and so on.}
\end{center}

Hagedorn asked how many different resonances of a state of given mass $M$ 
could contain or could decay into, given self-similar composition,
and he found the answer: exponentiallly many, $n(M) \sim \exp\{bM\}$. 
It is essentially a partition problem \cite{BFS} -- how many ways
are there of dividing an integer into integers? Consider as an illustration
the simplest case, ordered partitions. We then have

\newpage

$$
2=1+1,~2~  \rightarrow \rho(2)=2^1 
$$
$$
3=1+1+1,~1+2,~2+1,~3~  \rightarrow \rho(3)=2^2 
$$\be
4=1+1+1+1,~1+1+2,~1+2+1,~2+1+1,~1+3,~3+1,~2+2,~4~ \rightarrow \rho(4)=2^3
\ee

and hence in general

\be
\rho(n) = 2^{n-1} = (1/2) \exp\{n \ln 2\}
\ee

showing that the number of ordered partitions of an integer $n$ grows
exponentially with $n$. And such a growth in fact also holds in the 
more general case of self-similar resonance decay or composition.

\medskip

Hagedorn then went on to ask what effect such a resonance spectrum might have
on the thermodynamics of strongly interacting matter, and found a very
striking answer. Rather basic arguments \cite{BU,DMB} suggest that a
medium of {\sl interacting elementary constituents} (pions and nucleons)
could be replaced by an {\sl ideal gas of all possible resonance states}
arising from the interaction. 
That leads to the grand partition function
\be
\ln Z(T) \sim \int dM ~n(m)~\exp\{-M/T\},
\ee
which is seen to exist only if $T \leq 1/b$. In other words, there is an 
ultimate temperature 
of strongly interacting matter, the Hagedorn temperature $T_H=1/b$. For a 
while, Hagedorn took it to be the upper limit of the temperature of all
matter, just like there is a lower limit of -271$^{\circ}$ C. But a few years 
later, Cabbibo and Parisi \cite{CP} pointed out that it was the power $a$ of 
the prefactor in $n(M)=M^a\exp\{bM\}$ which determined if $\ln~Z(T)$ itself 
or only 
derivatives above a certain order would diverge at $T=T_H$. And such 
divergences are well known in statistical physics: they just correspond to
critical behavior. So $T_H$ only defined the end of the {\sl hadronic} state 
of strongly interacting matter; for $T > T_H$, there were no more hadrons, 
the matter had undergone a {\sl phase transition} into a new state. 
Today then, Hagedorn's limiting temperature is for us the first indication
of a new, hotter and denser state of strongly interacting matter. Once
hadron dynamics is correctly taken care of, hadronic thermodynamics defines 
its own limit, even without knowing anything about the quark infrastructure. 

\medskip
 
And Hagedorn's vision of the self-similar origin of the critical behavior 
turned
out to be fruitful far beyond expectation. Shortly afterwards, and quite
independently, Fortuijn and Kasteleyn \cite{FK1,FK2} showed that spontaneous 
symmetry
breaking as the origin of critical behavior in spin systems could equivalently 
be replaced by the fusion of clusters of all sizes. And still a little later,
also the dual resonance model \cite{DRM1,DRM2} was found to result in an 
exponential 
growth of the number of states, derived from the assumption of a complete
resonance determination of the scattering amplitude. This model subsequently 
became the progenitor of string theory, and even there the Hagedorn temperature
is still alive and well today.

\section{\large The Conjecture of Lucretius: Quark Confinement}

\vskip0.3cm
 
The other outcome of the multitude of resonances was their origin as composite
states of something more elementary. The different quantum numbers -- spin,
parity, charge, isospin, baryon number, and more -- ruled out the possibility 
of simply additive subconstituents and led to the quark model with its
non-Abelian composition laws. And this in turn provided the basis for 
what we now believe is the theory of strong interactions, {\sl quantum 
chromodynamics}. 
  
\medskip

We had already mentioned that reductionism, the idea that the complexity of
the visible world must arise from a few simple microscopic building blocks on 
an invisible level, was already introduced in ancient Greece. The perhaps most 
striking consequence in their logical derivation was put forward by the Roman
philosopher Lucretius \cite{Lucretius}, who noted around 50 B.~C. 

\medskip

{\sl So there must be an ultimate limit to bodies, beyond perception
by our senses. This limit is without parts, is the smallest possible
thing.  It can never exist by itself, but only as primordial part of
a larger body, from which no force can tear it loose.}

\medskip

I believe that this is the earliest reference to {\sl quark confinement}; 
with QCD, Lucretius' conjecture was finally, and for the first time, 
realized in the framework of physics.
Also the other essential result of QCD, {\sl asymptotic freedom}, was a novel
feature, distinguishing it from all previous composition schemes. The theory
had two basic pillars of support. The quark infrastructure now accounted for 
all
the observed resonances as bound states of different quantum number 
combinations. And for short distance physics, asymptotic freedom allowed
a perturbative approach, which provided a wealth of subsequently 
confirmed predictions. In spite of this immense success, three important
areas of strong interaction physics remained out in the cold: hadronic 
collisions in the non-perturbative regime (which meant more than 95 \% of 
all collision data), quantitative hadron spectra (calculate the masses of the
mesons and the baryons), and the thermodynamics of strongly interacting 
matter in the region of the transition from hadronic to quark matter.

\medskip

While high energy collisions in the non-pertubative regime still remain a
largely unsolved realm of QCD, a new avenue was found for the calculation
of hadron spectra and for the thermodynamics of strongly interacting matter.
It had been shown that the partition function, normally written in the 
canonical form with $\beta=1/T$,
\be
Z(\beta,V) = {\rm Tr}\exp \{-\beta {\cal H}\},
\label{part1}
\ee
could in field theory be reformulated as a Euclidean path integral
\cite{Bernard},
\be
Z_E(\beta,V) = \int ~{\cal D}A~{\cal D}\psi~{\cal D}{\bar\psi}~
\exp~\{-\int_V d^3x \int_0^\beta d\tau~
{\cal L}(A,\psi,{\bar\psi})~\}, 
\label{part2}
\ee
where $\cal H$ and $\cal L$ denote the QCD Hamiltonian and Lagrangian,
respectively. In the reformulation,
the trace over all particle states is replaced by path integrals
over the gauge ($A$) and quark/antiquark ($\psi/\bar \psi$) fields.
In this form, the integrations over space and temperature could be
discretized, i.e., replaced by lattice sums \cite{Wilson}. If one now 
carried out 
the integrations over the quark spinor fields and changed variables
from gauge fields to the corresponding $SU(3)$ color matrices
the structure of the resulting lattice partition function turned out 
to have the same form as that of a spin system on a lattice. And such
spin systems could be addressed by numerical simulations on high performance
computers \cite{Creutz}.

\section{\large Shift of Paradigm:\\
The Computer Simulation of Lattice QCD}

\vskip0.3cm

This not only provided a method to address strong interaction thermodynamics;
that was just one application of a new way of doing physics calculations, 
replacing for certain classes of problems analytical mathematics (which was 
not able to solve them) by numerical computer simulation. It has been 
called ``a betrayal of theoretical physics'', but, like many heresies, it 
opened new vistas and allowed in particular the quantitative study of many
aspects of collective behavior.

\medskip

In QCD, one used the exponential of the QCD action, $\exp\{-S(U,\beta)\}$, 
as weight to determine, essentially by throwing dice, equilibrium 
configurations for the state of the system. Given these, one could then 
``measure'' on them the observables of interest: energy density, pressure, 
specific heat, and more. In a way, it was a realization of Boltzmann's dream: 
one could now indeed generate for a many-body system all the possible
configurations allowed by the overall dynamics, energy and volume constraints,
assign them equal {\sl a priori} weights, measure the value of the
desired observable for each configuration, and then average. 

\medskip

The main drawback was that this method gave answers, but did not specify the
underlying reasons for the answers. Thus it was found that the energy density, 
as function of the temperature, suddenly increased in a very narrow region 
from rather low hadronic values to much higher values, not far from the 
Stefan-Boltzmann limit expected for an ideal gas of colored constituents 
\cite{EKMS1,EKMS2}. So the sudden increase seemed to be 
the onset of deconfinement. In pure gauge theory, strong interaction 
thermodynamics had in fact been initiated \cite{McL-S1,McL-S2,KPS}
by measuring the  expectation 
value $L(T)$ of the Polyakov,
\be
L(T) \sim \exp\{-F_Q(T)/T\},
\label{polya1}
\ee
and showing that it acted as confinement/deconfinement order parameter,
much like the magnetization did for spin systems.
Here $F_Q(T)$ is the free energy needed to separate a static quark-antiquark
pair infinitely far, and in the confinement regime, this diverged,
\be
F_Q(T) = \lim_{r\to \infty} F_Q(r,T) \to \infty,
\label{polya2}
\ee
causing $L(T)$ to vanish there. In a deconfined medium, in contrast,
gluon screening left $F_Q(T)$ and hence also $L(T)$ finite.

\medskip

But in full QCD, quarks acted like an external field, leading to some
alignment of the generalized spin $L(T)$ at all temperatures, just as
the magnetization in the Ising model always remains finite for a 
finite external field. So how should one now specify deconfinement?
In addition, there was the ``other'' transition, the restoration of chiral
symmetry. At low temperatures, gluon dressing gave the quarks a constituent
mass of hadronic scale; with increasing temperature, this dressing would
melt, shifting the effective quark mass back to its (almost vanishing)
current quark value. Did this mass shift coincide with color deconfinement? 
In the chiral limit, for vanishing quark masses, the chiral condensate was 
indeed a {\sl bona fide} order parameter, but what about deconfinement?

\medskip

Today, we may have at least a partial answer. Consider QCD with two massless
quark flavors. The chiral condensate identifies the chiral transition 
temperature $T_{ch}$. One way of defining deconfinement is to consider
the transition at which the average baryon number per constituent flips from
unity (in the hadronic phase, with nucleons) to 1/3, in the quark phase.
This is evidently the point at which quarks are no longer bound to color
neutral three-quark entities. And it is indeed observed in lattice studies,
for higher moment fluctuations, that this point coincides with the
chiral symmetry restoration temperature. So in what one could call an
{\sl ab initio} approach, we conclude that deconfinement and chiral 
symmetray restoration coincide, occurring (according to latest results)
at a unique critical temperature of about 160 $\pm$ 10 MeV.

\medskip

But what these calculations cannot tell us is what happens just beyond 
the transition. It is clear, and has been for twenty years, that the
plasma of deconfined color is strongly interacting. The crucial measure,
the so-called trace anomaly
\be
\Delta(T) = {\e - 3P \over T^4}
\label{delta}
\ee
comes from small values to reach a peak just above, but definitely
above, the critical temperature. Beyond this peak, it decreases, but
it reaches something like perturbative behavior only at very much
higher temperatures, if at all.

\medskip

\begin{figure}[h]
\centerline{\psfig{file=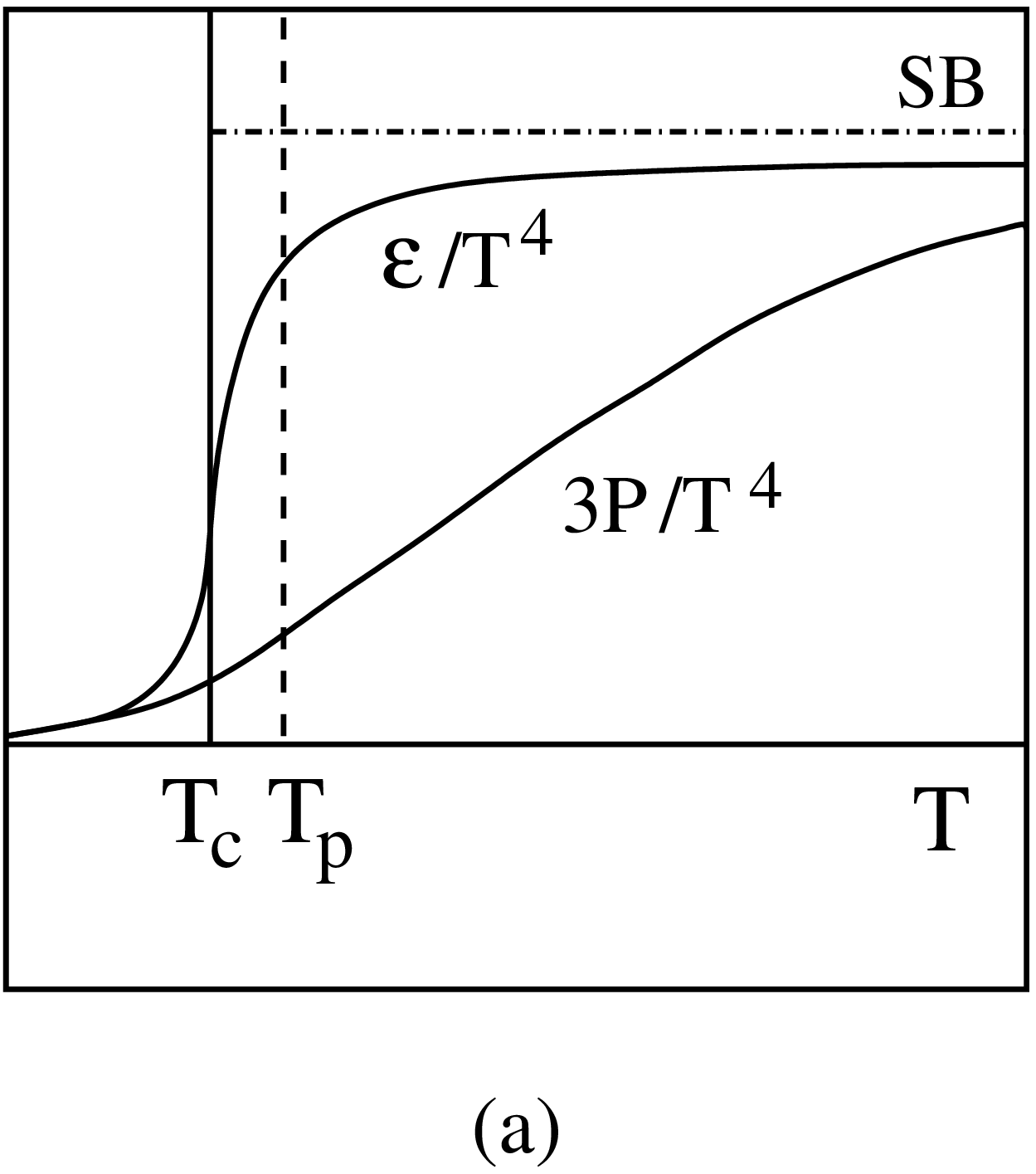,width=4.17cm}\hskip1.5cm
\psfig{file=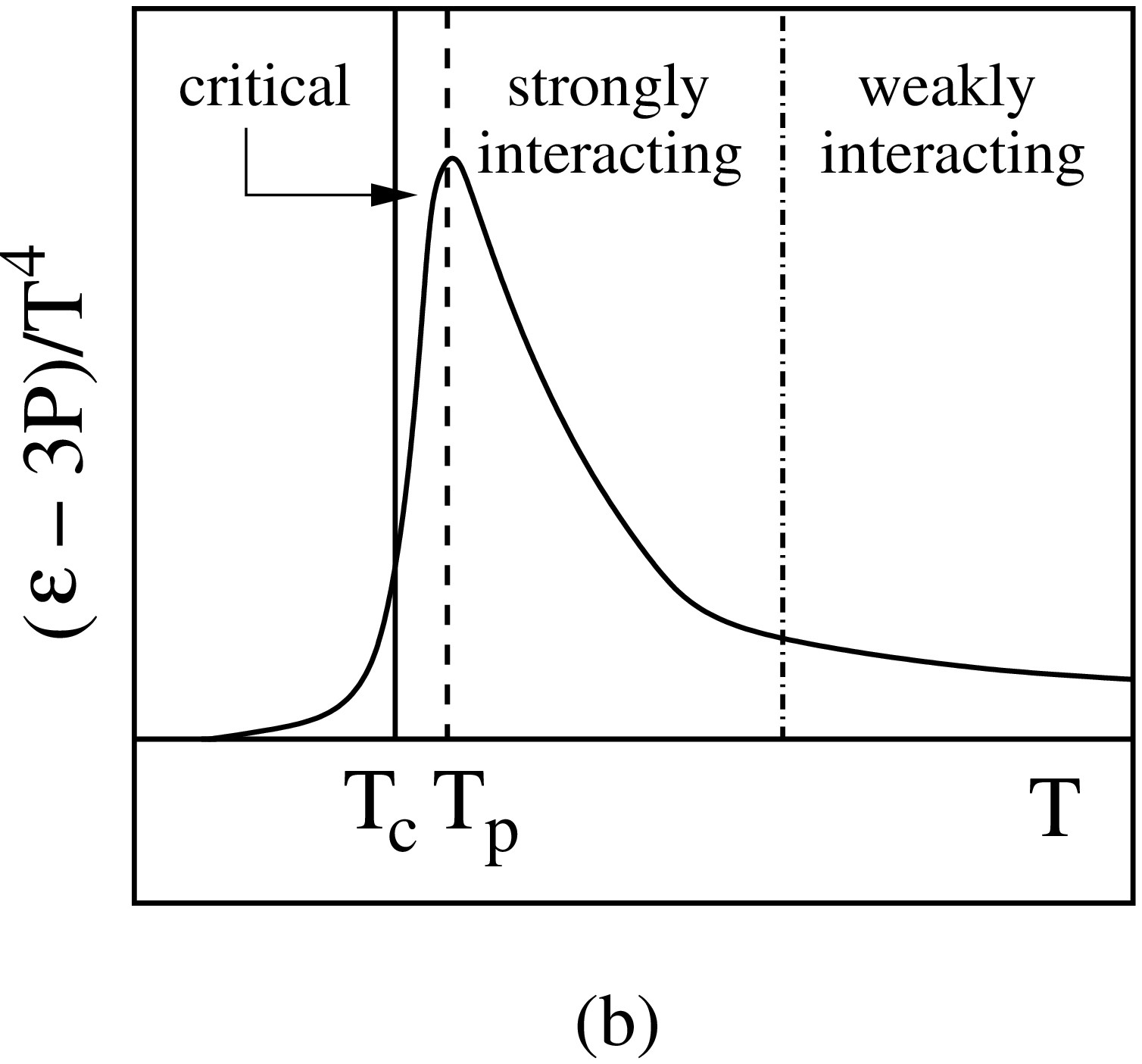,width=5.1cm}}
\caption{Schematic view of energy density and pressure (a) and of 
the interaction measure (b) in the deconfined medium; in (a), SB denotes
the Stefan-Boltzmann limit.}
\label{sQCD}
\end{figure}

So we can summarize our present state of knowledge in statistical QCD,
as obtained from lattice studies, schematically as shown in Fig.\ \ref{sQCD}.
Approaching the transition temperature $T_c$, initially the energy density 
increases faster than the pressure, as a reflection of critical behavior. 
At a certain temperature $T_p > T_c$, the two roles are interchanged, since 
eventually the system tends towards the Stefan-Boltzmann limit 
$\e/T^4 \to 3P/T^4$.
The peak of the interaction measure is the result of the interchange.

Here again we note both the advantages and the problems of the lattice
simulation. We obtain beautifully the over-all pattern of the crucial
thermodynamic observables, but we don't really know the specific 
``microscopic''
or ``dynamic'' origins of this behavior. On the confined hadronic side,
the model of an ideal resonance gas, as first proposed by Hagedorn and since
then refined to become a very useful tool of analysis, provides us with
a fairly detailed understanding of what is happening; we return to this
in section 6. On the plasma side,
there exist various models, but one probably has to admit that the origin
of the observed behavior is not really understood.

\medskip

A serious shortcoming of the computer simulation approach is its (present?)
failure for systems of large baryon density. The origin of the difficulty
seems purely technical: the weight-function used to generate equilibrium
configurations is no longer positive definite at non-zero baryonic potential
$\mu$. In other words, the integrals to be evaluated now have integrands
fluctuating between positive and negative values. All that is known (or
believed to be known) for the QCD phase diagram as function of temperature 
$T$ and baryonic potential $\mu$ is thus based on approximate lattice 
methods, trying extensions from vanishing to finite $\mu$, or on 
more or less phenomenological models.  

\medskip

\begin{figure}[htb]
\centerline{\psfig{file=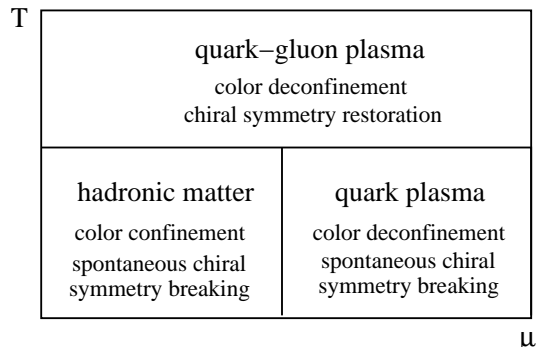,width=7cm}}
\caption{Schematic view of a possible QCD phase diagram}
\label{Qplasma}
\end{figure}

My own favorite model assumes color deconfinement to set in 
when the density of 
constituents surpasses the hadronic scale, i.e., when each quark sees
within its confinement range several other quarks, and chiral symmetry
restoration when the temperature of the medium melts the gluon dressing
of the constituent quarks. At $\mu=0$, the two coincide, since the 
density increase is due to the temperature increase. For $T=0$, that is
not the case, and so one expects deconfinement there to lead to a 
plasma of massive quarks. The resulting phase diagram is schematically
illustrated in Fig.\ \ref{Qplasma}.

It should be noted that the deconfined quarks could, at low temperature,
bind to form bosonic colored diquarks, and these in turn could condense
to form a color superconductor. Such a phase, as a special form within the
quark plasma,  is not included in Fig.\ \ref{Qplasma}.

\section{\large The Little Bang: Making Matter in Collision}

\vskip0.3cm

Given what we know about strong interaction thermodynamics, how can we
apply it? Neutron stars are far away, and the Big Bang was long ago.
Is there some way to create strongly interacting matter in terrestrial
experiments? We had already mentioned heavy ion collisions, and in the 
course of the 1980's, that possibility gained more and more interest.

\medskip

The idea is quite straightforward. At the start of the program, beams of 
energetic nuclei were made to hit stationary nuclear targets. In the modern 
versions, however, at the LHC/CERN as well as at the RHIC/BNL, two beams of 
nuclei, moving 
in opposite directions, collide ``head-on'' in the detectors of the 
experiment. Through Lorentz contraction in the longitudinal direction, the 
incoming nuclei appear as pancakes to an observer in the laboratory.
In the collision, they overlap and form a system of many quarks contained 
in a small disk, an excited colored medium. This highly compressed bubble 
subsequently expands, still retaining its colored nature, and through
many interactions it could form a thermal system: a quark-gluon plasma.
This colored plasma would then cool off and lead to the production of the 
hadrons detected in the laboratory.

\medskip

From the very beginning,
T.\ D.\ Lee was one of the main proponents of such a new research program,
and he moreover explained the idea to the famous Chinese painter Li Keran, 
who in 1989 composed a picture of two fighting bulls, with the heading

\begin{figure}[h]
\centerline{\epsfig{file=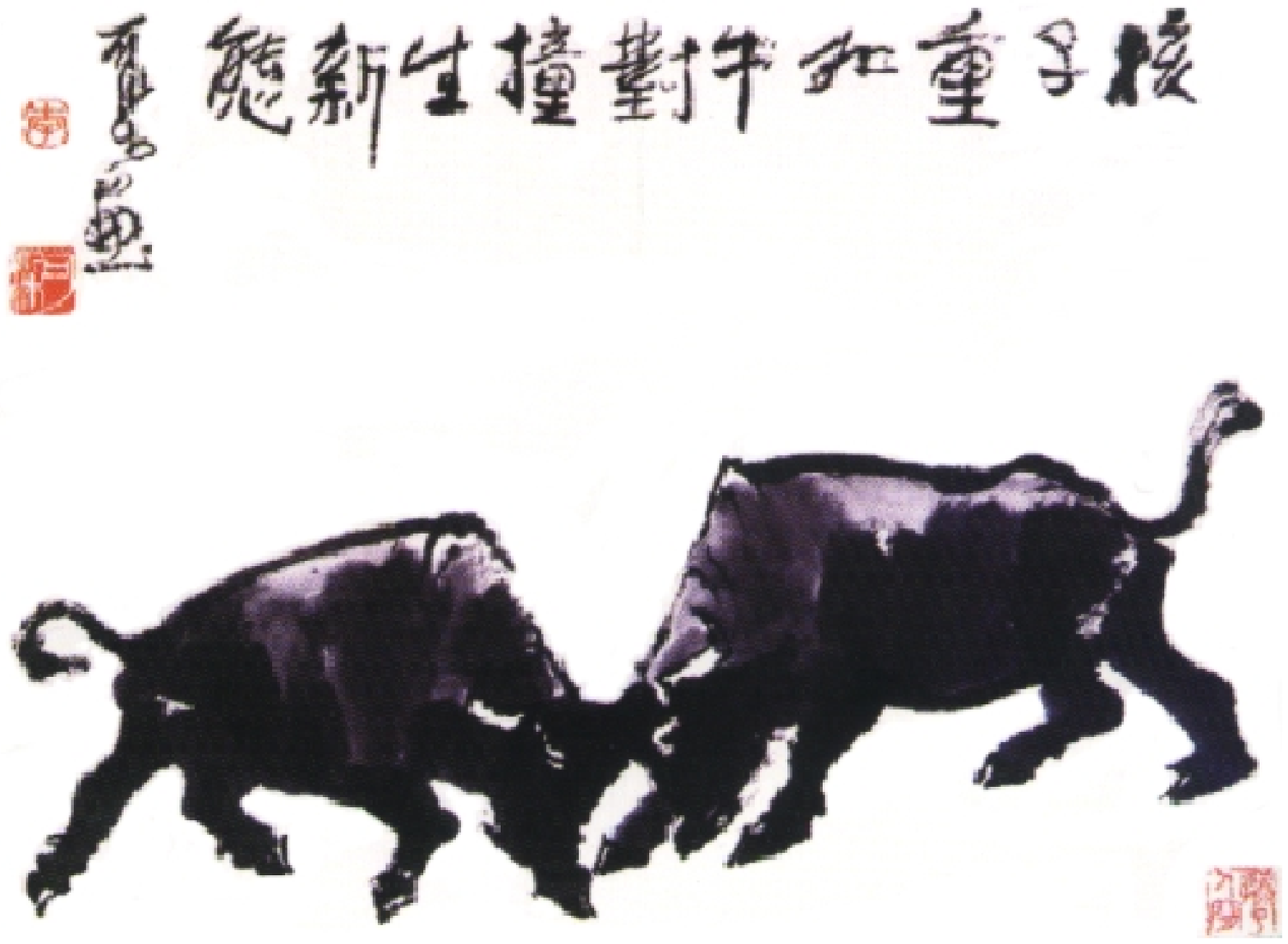,width=6cm,angle=-90}}
\end{figure}

\centerline{\sl Nuclei as heavy as bulls through collision
generate new states of matter.}

\bigskip

Today the bulls also exist as a beautiful life-size sculpture, close to the 
campus of Tsinghua University in Beijing, as probably the only existing
memorial to heavy ion collisions. The justification for a research 
program devoted to the empirical study of quark matter is quite different 
from most others in recent times. It is not the search for a well-defined 
and theoretically predicted entity, such as the Higgs' boson, ultimate aim 
of several presently ongoing experiments at CERN and Fermilab. It is also not 
the purely exploratory study of the strong interaction in the last century: 
what happens if we collide two protons at ever higher energy? Instead, it is 
almost alchemistic in nature: can we find a way to make gold? Is it possible, 
through the collisions of two heavy nuclei at high enough energy, to make 
strongly interacting matter of high energy density and study its behavior
in the laboratory?

\medskip

The opinions on the feasibility of such an endeavor were mixed. Richard 
Feynman was as always ready with a concise judgement and said ``if I throw 
my watch against the wall, I get a broken watch, not a new state of matter''. 
Actually, the problem had two sides. One was the aspect Feynman had 
addressed: is it possible to create through collisions something we would 
call matter? The other was the question of whether the experimentalists would 
be able to handle the large number of hadrons to be produced in such 
collisions. Both were indeed serious, but the promise of finding a way to 
carry out an experimental study of the stuff that made 
up the primordial universe -- that promise was enough to get a first
experimental program going, in 1986, at Brookhaven and at CERN.
To minimize the costs, both Labs used existing injectors, existing 
accelerators (BNL-AGS, CERN-SPS), existing detectors and, 
as someone pointed out, existing physicists not needing additional 
salaries: one big recycling project. Whatever, the second of
the two questions mentioned above was indeed and resoundingly answered
in the affirmative. At present collision energies, one single
interaction of two heavy nuclei produces some thousands of new particles
(see Fig.\ \ref{RHIC}), and the detectors, the analysis programs and the 
experienced physicists can handle even that. 

\medskip

The first and conceptually more serious question is not yet answered
as clearly, and in the physics community at large (in contrast to the
heavy ion community), there are still adherents of Feynman's point of view. 
It is clear that present collisions, with a center of mass energy of more 
than a million GeV for a central lead-lead interaction at the LHC, provide 
the highest energies ever reached on Earth, and the interactions produce 
thousands 
of new particles in a rather small volume, so that also the initial density 
of constituents is extremely high. The canonical estimate for the energy
density produced in a central $A-A$ collision is given by the 
Bjorken form \cite{Bjo}
\be
\e = {\bar p_0 \over \pi R_A^2 \tau_0} \left({dN_h \over dy}\right)_A,
\label{bj1}
\ee
where $(dN_h/dy)_A$ is the number of hadrons produced in a unit of 
central rapidity, $\bar p_0$ the average energy per hadron, $R_A$ the nuclear 
radius and $\tau_0 \simeq 1$ fm the equilibration time needed to produce a 
thermal medium. For gold-gold collisions at RHIC (top energy) this leads
to some 6 - 7 GeV/fm$^3$, and for lead-lead collisions at the LHC to 
10 GeV/fm$^3$ or more. These are values exceeding the energy density
within a single hadron by up to a factor twenty. So the early medium was
indeed hot, and it seems possible to understand what is happening in this
stage only in terms of quarks, gluons and their interactions. But does that 
allow us to speak of quark {\sl matter}? What are the essential features of 
matter? And how can one show that the media produced in high energy nuclear 
collisions share these features, both in the early stages and in the later 
evolution? 

\section{\large The Abundance of the Species:\\ 
Universal Hadrosynthesis}

\vskip0.3cm

Multiparticle production in high energy collisions of strongly interacting
particles has fascinated physicists for well over half a century. The little 
bang of such collisions produces with increasing energy an ever growing number 
of mesons and baryons of an ever growing number of species, and from the 
beginning, these large numbers were a challenge to describe the reactions 
by collective or statistical approaches. It was tempting to go
even further and imagine that the collisions really produced 
droplets of strongly interacting matter, thus providing a means 
to access in the laboratory the thermodynamics of strong interaction 
physics.

\medskip

The main features observed in high energy collisions are the multiplicity, 
i.e., the number of produced particles as function of the collision energy,
the momentum spectra of the particles, and the relative abundances of
the different species. And the detailed study of the abundances has led to
what seems to me the perhaps most striking result of high energy strong
interaction physics.

\medskip

We have seen that high energy collisions produce an initial state of an
energy density much higher than that of any hadronic medium. Leaving open
the issue of whether or not this hot medium in equilibrium, it
will expand, cool off and eventually hadronize. At this point, we have
a medium of strongly interacting hadrons, and as already argued by Hagedorn,
this system can be considered as an ideal gas of the possible resonances.
Using Hagedorn's self-similar resonance pattern, that had led to the 
prediction of the critical temperature $T_H$. But we can take a more
pragmatic point of view and simply calculate the partition function of
all the resonances listed in the Partical Data List \cite{PDG}, with all 
their known
discrete quantum number states. This will require
us to stop at some high mass, since resonances of masses above some 2 - 3
GeV are poorly known. But if the effective temperature of the medium is
not too high, that should not matter too much; in any case, one can check
what a mass cut does to the resulting description. The partition function
of the medium is then, for vanishing baryon density, given by
{\be
\ln Z(T,\mu,V) = {V T \over 2 \pi^2} \left[\sum_i^{\rm mesons}d_i m_i^2 
K_2(m_i/T) + 2\cosh(\mu/T)\sum_i^{\rm baryons}d_i m_i^2 K_2(m_i/T)\right], 
\ee}
where $d_i$ denotes the degeneracy of the resonance (spin, charge, etc.),
and $\mu$ the baryonic chemical potential. Of course, charge and strangeness
conservation has to be taken into account; we have here just suppressed it
for the sake of simplicity.

\medskip

\begin{figure}[h]
\centerline{\psfig{file=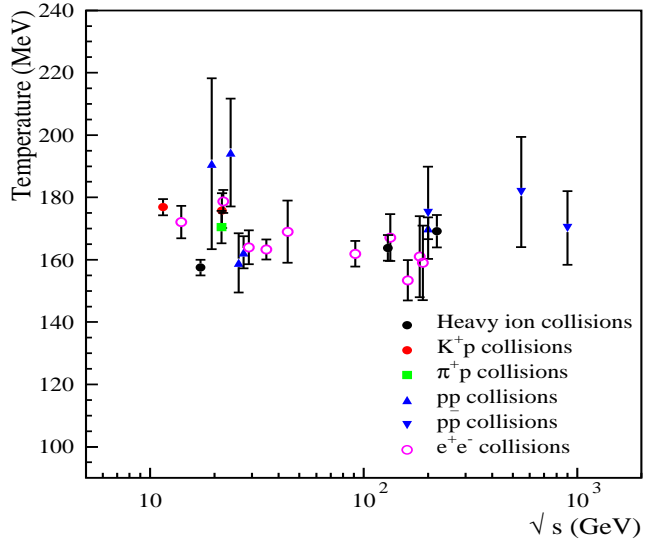,width=9.5cm,height=8cm}}
\caption{Hadronization temperatures obtained from resonance gas analyses
of hadron production from various initial
collision configurations at different (high) energies.}
\label{T-all}
\end{figure}

Considered as a function of temperature, this partition function will never
lead to critical behavior; it never diverges, for two reasons. The mass cut
makes it a power series in temperature, always convergent. But even if we
remove the mass cut, the empirical degeneracy factors $d_i$ are not
exponential and therefore cannot compensate the $\exp\{-m_i/T\}$ contained
in the Hankel functions $K_2(x)$.  Nevertheless, the ``shadow'' of the 
transition makes its appearence, as it turned out, if we consider the
abundances of different species $i$ and $j$. They are (at $\mu=0$) given by
\be
{N_i \over N_j} \simeq \left({d_i m_i  \over d_j m_j}\right)^2 {K_2(m_i/T_H) 
\over K_2(m_j/T_H)} \simeq \left( {d_i m_i \over d_j m_j}\right)^{3/2}\!
\exp\{-(m_i - m_j)/T_H\}. 
\label{abun}
\ee
And the curious thing, nature's hint, was that apart from minor deviations,
the observed abundances of {\sl all species} in {\sl all high energy 
collisions}, whether in $e^+e^-$ annihilation, in $p-p$ or in nuclear 
collisions, always pointed to the same hadronization temperature 
\cite{RGM1,RGM2}. Up to thirty different
hadron species all agreed in their abundances that they were formed at
a universal temperature of about 160 -- 170 MeV. A recent summary compilation
is shown in Fig.\ \ref{T-all} \cite{RGM1,beca-sum}. 
Just as the cosmic background radiation was 
left over from the Big Bang at the time of last scattering, at the end of the 
radiation era, at $T_{\rm rad} \sim 3000^{\circ}$ K, so all observed hadrons 
were
formed at what statistical QCD had obtained for the hadronization temperature
$T_H \simeq 170 \pm 10$ MeV of a quark-gluon plasma. 

What does this tell us? Here opinions in the community differ. It is obviously
tempting to consider the many partons of two colliding nuclei to interact and
thermalize, and to have the resulting QGP then hadronize, just like vapor
condenses below 100$^{\circ}$C. In other words, the thermal nature of the
hadronic state is taken as a result of a prior thermal quark-gluon state,
which in turn was formed through thermalization of the (non-thermal) initial
multiparton state.
This leaves open the question of why $e^+e^-$ annihilation, in two-jet
events just one hard quark-antiquark pair, and similarly $p-p$ interactions,
lead to the same form of thermal behavior. I don't believe that one should
ignore such clear and pointed hints of nature. There must be a universal 
hadronization mechanism, operative whenever a quark and an antiquark reach 
their confinement horizon\cite{CKS}; this can happen in jet cascades as well 
as in a cooling QGP. 

\section{\large Melting of Quarkonia: The QGP Temperature} 

\vskip0.3cm

In the search for probes of the early hot stages of heavy ion collisions,
measuring the temperature of stellar cores can serve as an inspiration. This 
temperature is determined largely through a spectral analysis of the light 
the stars emit. Their interior is generally so hot that it is a plasma of 
electrons and nuclei, emitting a continuous spectrum of light, and the 
frequency of this light is proportional to the energy density of the inner 
core. In the cooler outer corona of the star, atoms can survive, and the 
passing light from the interior excites their electrons from the ground 
state to higher level orbits. The photons doing this are thereby removed 
from the continuous spectrum, and this shows up: there are absorption lines, 
whose position indicates the elements present in the corona, and whose 
strength measures the energy of the light from the interior. To take the 
simplest case: if the corona contains hydrogen atoms, then the frequencies 
needed to excite these to the different excitation levels are candidates 
for absorption lines. In the case of relatively cool stars, the photons 
will not be energetic enough to do much except to bring the atoms into 
their lowest excited state. Sufficiently hot stars, on the other hand, 
will generally result in jumps to higher excitation states. So by looking 
which excited states are the target of the photons, we can tell what the 
temperature of the stellar core is.

\medskip

\begin{figure}[h]
\centerline{\psfig{file=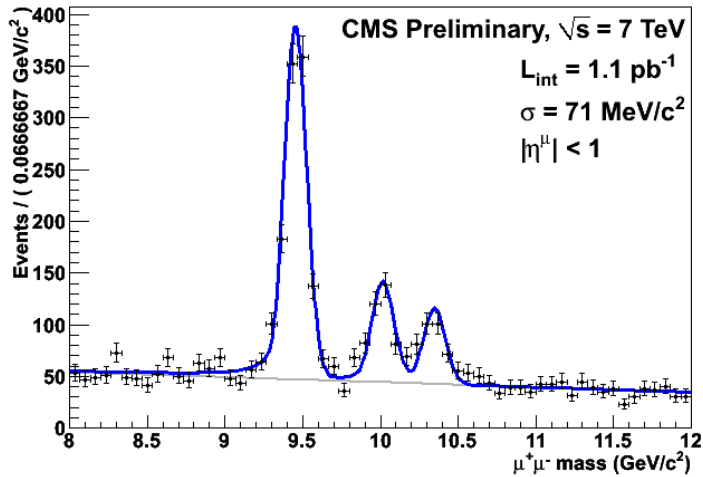,width=7cm,angle=-90}}
\caption{The dimuon spectrum in the bottomonium range, measured in $p-p$ 
collisions at the LHC for a collision energy $\sqrt s = 7$ TeV. The peaks 
correspond to the $\Upsilon(1S),~\!\Upsilon(2S)$ and $\Upsilon(3S)$ states,
respectively.}
\label{upsi1}
\end{figure}

A similar method can be applied to study the early interior of the
medium produced in high energy nuclear collisions. Here one can observe
the mass spectrum of dileptons, replacing the photon spectrum from 
stars. Ideally, this dilepton spectrum arises from the annihilation of
quark-antiquarks pairs in the hot plasma; in practice, a number of competing
sources come into play and have to be eliminated. This has so far made 
the identification of the thermal radiation from quark matter rather 
difficult. But on top of the smooth curve found for the dilepton mass 
distribution there are  are some very pronounced sharp peaks at
well-defined positions (see Fig.\ \ref{upsi1}): they are the signals of 
quarkonium production in nuclear collisions, and they can play the role 
which the atoms in the corona had in the stellar specral analysis.

\medskip

Quarkonia are unusual hadrons, bound states of the heavy charm ($m_c \simeq
1.3$ GeV) and bottom quarks ($m_b\simeq 4.5$ GeV) and their antiquarks. They 
are very much smaller
than normal hadrons ($r_{\Upsilon} \simeq 0.1$ fm), and their ground states 
are very tightly bound ($\Delta E_{\Upsilon}= 2M_B - M_{\Upsilon} > 1$ GeV).
That's why both charmonia and bottomonia can in principle survive
deconfinement and exist as color-neutral hadrons in a colored
quark-gluon plasma,
up to some temperature. The different binding energies of the different
quarkonia imply that they melt in a hot QGP at different temperatures,
and the quarkonium spectrum can therefore serve as plasma thermometer 
\cite{MS,KMS,KS}.

\medskip

That leaves theory with the charge of calculating the dissociation points
of quarkonia in a hot deconfined medium, and experiment with the problem of
determining the parameters (collision energy, energy density, temperature) 
at which the quarkonia disappear in the dimuon spectrum emitted in nuclear
collisions \cite{KluSa}. If both problems were solved, and if the numbers 
would match,
we would have quantitative evidence for the production of a quark-gluon
plasma in heavy ion collisions. 

\medskip

The calculation of the in-medium dissociation temperatures of quarkonia
has been pursued for over twenty years, and we are still waiting for the
final word. Initially, one made a model for the screened heavy quark
potential in the QGP \cite{KMS,KS}. When lattice data for the heavy quark
interaction became available, that was employed in numerous studies
to determine the potential \cite{HS-rev}. It led to a still not resolved
issue: should one use the free energy or the internal energy of the 
$Q\bar Q$ for the potential? At $T=0$, the two coincide, but for finite $T$,
the internal energy leads to a stronger binding. Arguments based on QED
indicate that there the free energy is the relevant quantity
\cite{Laine1,Laine2,Blai}. But
in the weak-coupling regime the binding structure is very different
from the strong-coupling QCD world, since now the separation of the $Q$
and the $\bar Q$ requires both the work against the binding force and
the energy to create the gluon dressing of the separated quarks. If 
both contribute to the binding, the $J/\psi$ will survive up to about $2~T_c$;
the free energy alone lets it melt around $1.2~T_c$. -- Another open issue
is width of the quarkonium states, and related here, the imaginary part of 
the potential.

\medskip

The solution will eventually come from direct lattice calculations,
which are in fact in progess, thanks to the availability of ever more
powerful computing facilities. They provide an integral transform of
the desired in-medium quarkonium spectrum (the so-called correlator),
and the inversion of this transform, given the precision of the 
calculations and probabilistic inversion technique (MEM: maximum
entropy method), is not yet unambiguous. Present results seem to
lie in the range of 1.5 - 2.0 $T_c$ for the dissociation of the 
$J/\psi(1S)$.

\medskip

\begin{figure}[htb]
\vspace*{-0mm}
\centerline{\psfig{file=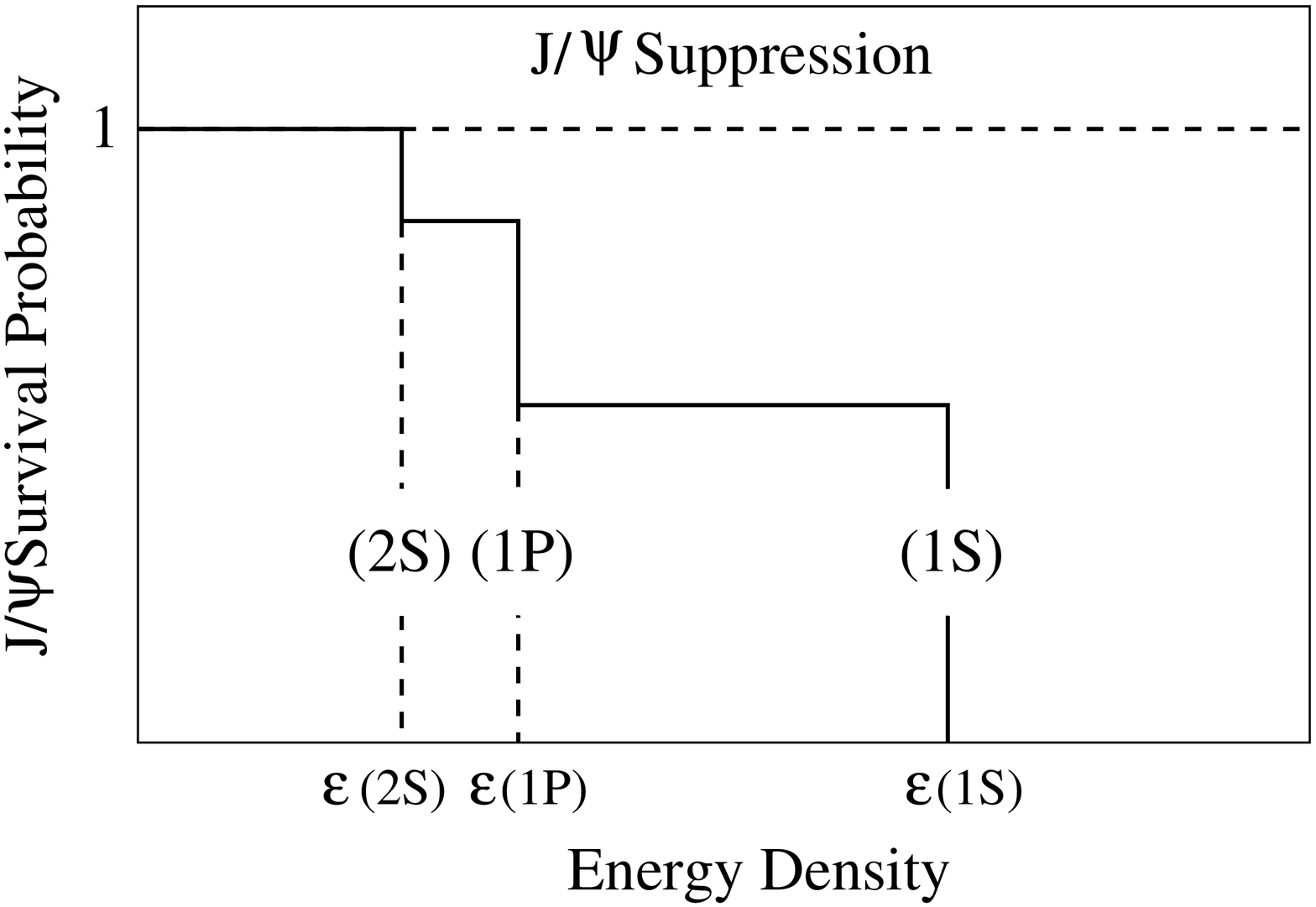,width=6.5cm,height=5cm}\hskip1cm
\psfig{file=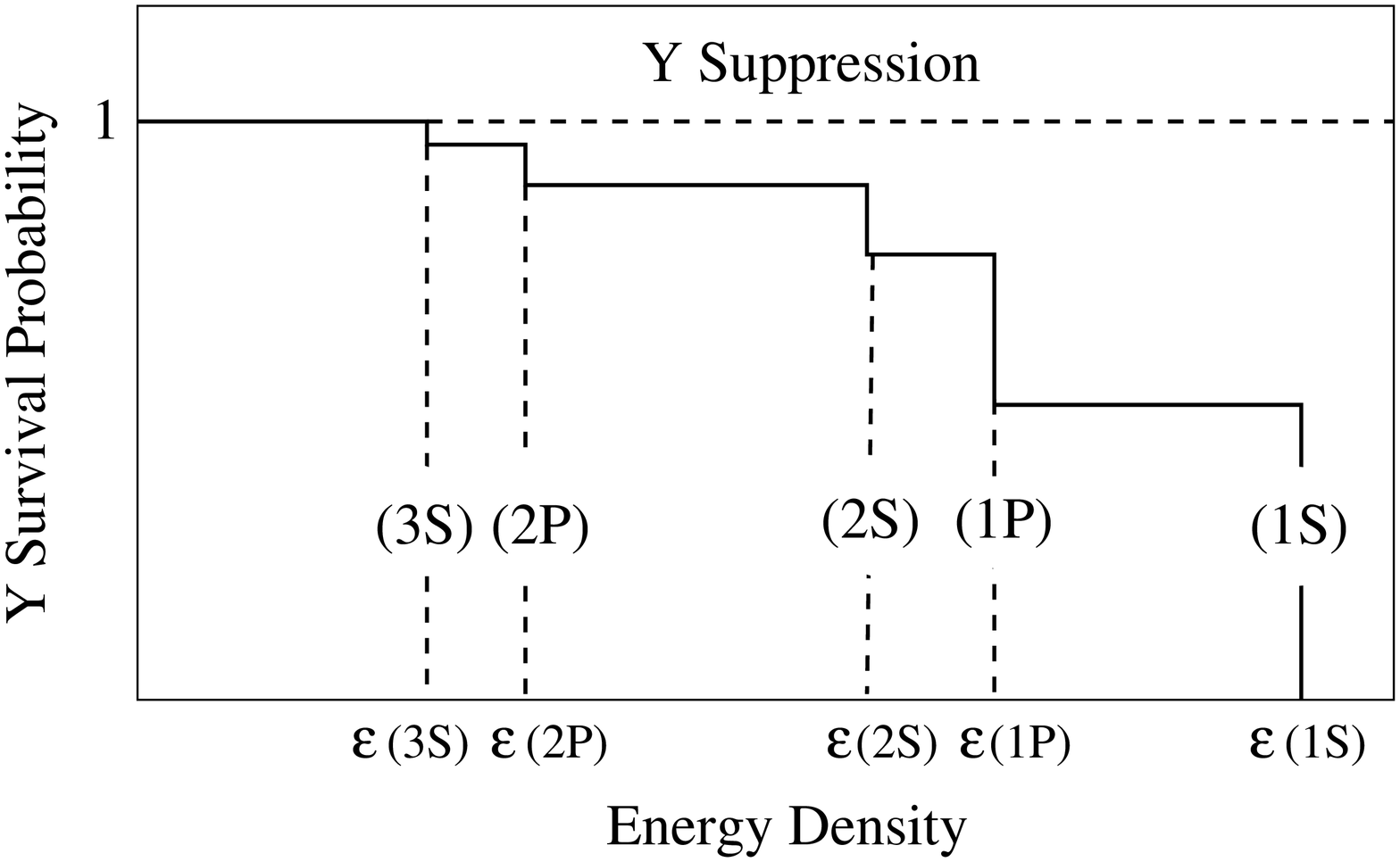,width=7cm,height=5cm}}
\caption{Sequential quarkonium suppression}
\label{seq}
\end{figure}

It should be emphasized that we need, in fact, the dissociation temperatures
of all the quarkonium states, not just those for the ground states. The
experimentally observed \J~and $\Upsilon$ are only in part directly produced;
almost half come from the decay of higher excited states. Since these
states are very narrow, the decay occurs far outside the interaction region,
and so the medium affects the passing excited states. As a result, one
predicts {\sl sequential quarkonium suppression} \cite{KKS}: first the
fractions from the excited states are suppressed, then eventually also the
ground states. The result is a stepwise suppression pattern as function 
of the energy density or temperature, see Fig.\ \ref{seq}.

\medskip

\medskip

On the experimental side, one also encounters a variety of complicating
factors.  Before the effect of any QGP on quarkonia can be determined,
the role of more profane initial and final state effects has to be brought
under control: shadowing/antishadowing, parton energy loss, nuclear and
hadronic absorption, etc. The elimination of these began at the SPS by
measuring \J~suppression relative to the Drell-Yan dimuon background.
Later, and at RHIC, this was replaced by combining \J~$p-p$ data with
nuclear effects determined in $p/d-A$ collisions. At the LHC, this is
also in progress. Nevertheless, one should keep in mind that all these
schemes are in a sense ``crutches''; the obvious and natural quantitity
to consider is the ratio of quarkonium production to that of open
charm or bottom. In this ratio, most initial state effects will cancel 
out, and it addresses directly the question of whether the fraction of
charm or bottom production going into the hidden sector is reduced or
fully suppressed by the presence of the QGP \cite{Sridhar}.

\medskip

Another interesting aspect has come up in the study of \J~production.
We have so far considered {\sl primary} quarkonium production, i.e.,
a $c \bar c$ pair is produced in a specific nucleon-nucleon collision
and subsequently, a certain small fraction combines to form a \J, the
remainder lead to open charm. If the medium would contain sufficiently 
many remaining unbound pairs, and if these would become part
of the thermalized plasma, then at the hadronization point, a $c$ from
one collision could statistically combine with a $\bar c$ from another
collision to make a \J \cite{PBM,Thews1,Thews2,Rapp}. Such {\sl secondary} 
\J~production would then result in an overall enhancement of the 
rate. Clearly, a necessary prerequisite is that there is a sufficiently
large number of available charm/anticharm quarks, i.e., the nuclear collision
energy has to be high enough. The prediction of such statistical combination
is just the opposite of sequential suppression: while the latter has the
production rate of hidden charm relative to open charm decrease or vanish
in nuclear collisions, compared to $p-p$ interactions, statistical combination
predicts an increase of this ratio. 

\medskip

A crucial input for the statistical combination mechanism is that indeed the
mere statistical presence of a $c$ and $\bar c$ suffices to form a \J. 
Additional dynamical conditions (sufficiently close approach, etc.),
could strongly affect the predictions. On the other hand, if statistical
combination is in effect, i.e., if the suppressed primary production is
more than compensated by the statistical secondary process, then this
would constitute evidence for a thermalization of even the heavy charm
quarks. It would remove, however, the possibility of checking experimentally
QCD predictions for charmonium dissociation.

\medskip

\begin{figure}[h]
\centerline{\psfig{file=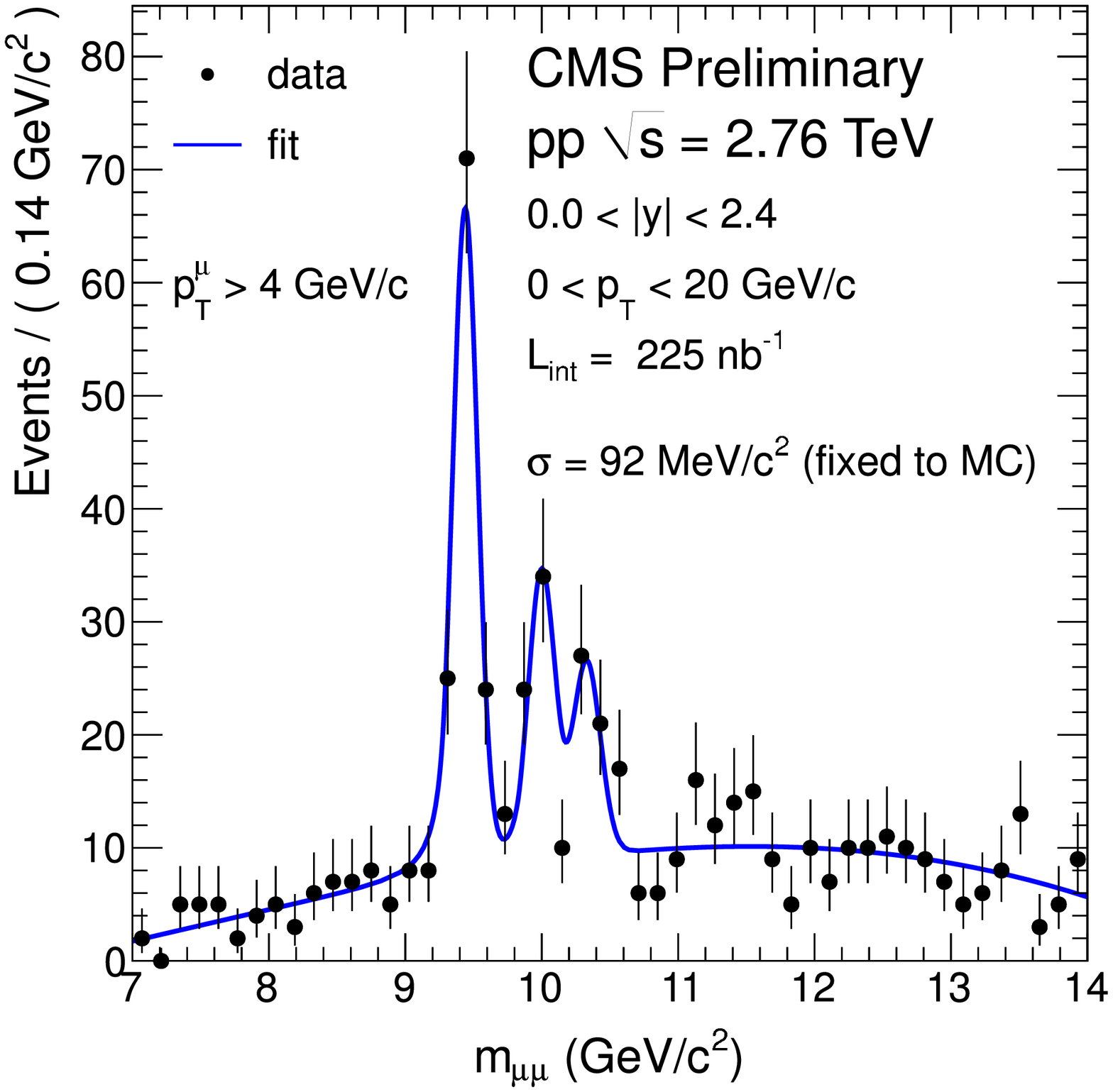,width=7cm}\hskip1.5cm
\psfig{file=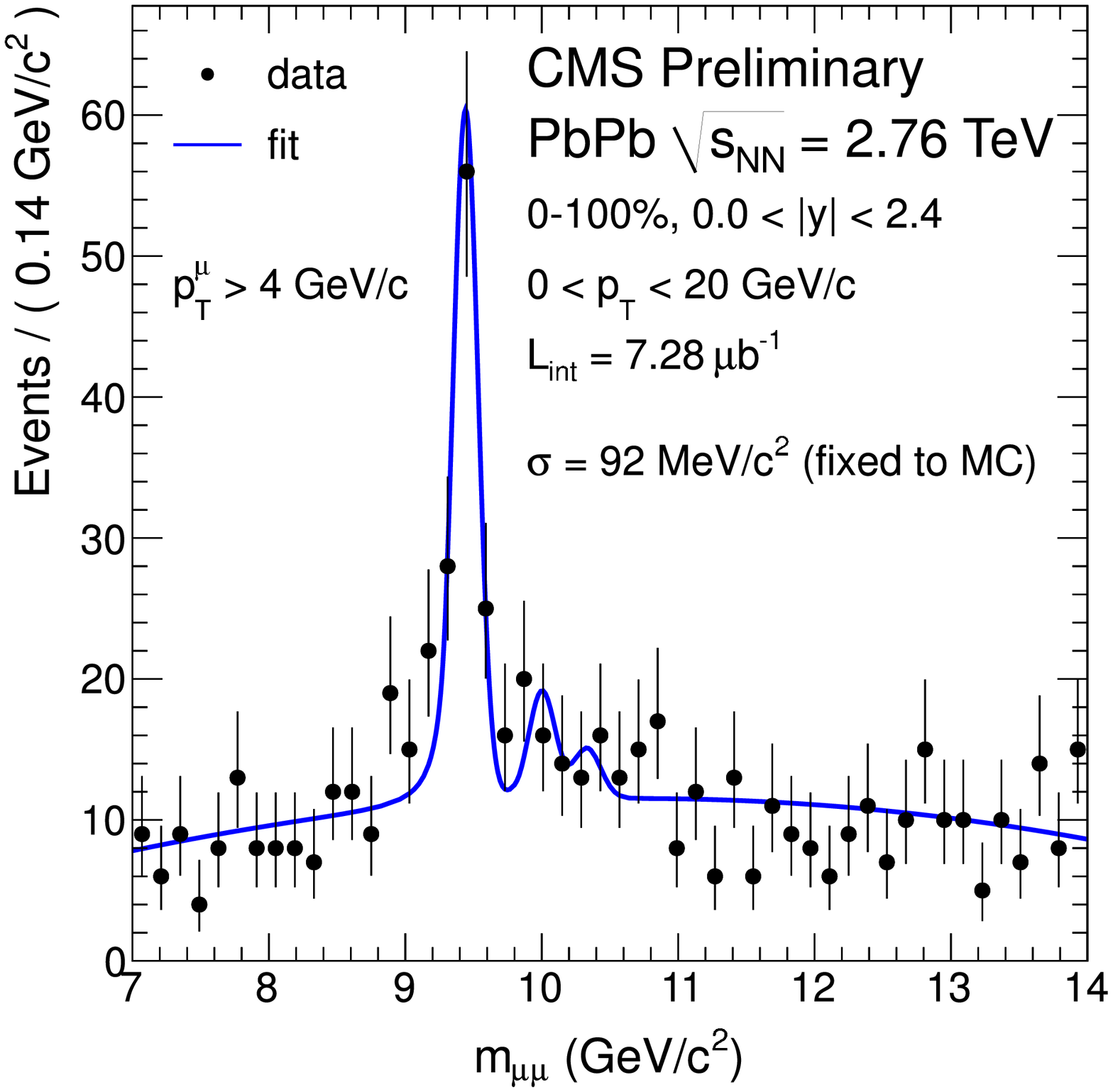,width=7cm}}
\hskip3.5cm (a) \hskip8cm (b)
\caption{The dimuon spectrum in the bottomonium range, measured (a) in $p-p$ 
and (b) in $Pb-Pb$ collisions at the LHC, both for a collision energy $\sqrt s 
= 2.76$ TeV. The peaks in (a) correspond to the $\Upsilon(1S),~\!\Upsilon(2S)$ 
and $\Upsilon(3S)$, respectively \cite{CMS-upsi}.}
\label{upsi2}
\end{figure}

This check would then have to be applied to bottomonium production, for
which even at LHC energies statistical modifications seem excluded. And
here on does indeed observe at $\sqrt s = 2.75$ TeV a suppression of the
higher excited state relative to the $\Upsilon(1S)$ \cite{CMS-upsi}, as
shown in Fig.\ \ref{upsi2}. Perhaps a more detailed quantitative study
of this suppression, both in amount and in onset thresholds, will allow
a direct comparison to statistical QCD calculations.    


\section{\large Quenching of Jets: The QGP Density}

Probing the density of the medium produced in the early stages of nuclear
collisions is also a task which finds parallels in other fields of physics. 
A standard tool for this is the study of the attenuation of a flux of 
fast particles losing energy by scattering in the course of their passage
through a medium.

\medskip

In our case, we consider the production of jets in nucleus-nucleus 
collisions, consisting of one or more hadrons with very high momentum 
transverse to the collision axis. These jets are formed initially 
by a very energetic parton, quark or gluon, produced in the early hard
collision stages and emitted in the transverse direction. Given a 
QGP formation time of about 1 fm, the nascent jet will pass through
several fermi of hot deconfined matter before it escapes the interaction
region and eventually hadronizes. How much energy it has lost when it 
finally emerges will tell us something about the density of the medium 
\cite{Bj,GW,B,Z}. In particular, the density in a quark-gluon plasma
is by an order of magnitude or more higher than that of a confined
hadronic medium, and so the energy loss of a fast passing color charge 
is expected to be correspondingly higher as well. Let us consider
this in more detail.

\medskip

An electric charge, passing through matter containing other bound or
unbound charges, loses energy by scattering. For charges of low incident
energy $E$, the energy loss is largely due to ionization of the target
matter. For sufficiently high energies, the incident charge scatters
directly on the charges in matter and as a result radiates photons of
average energy $\omega \sim E$. Per unit length of matter, the
`radiative' energy loss due to successive scatterings is thus proportional 
to the incident energy.
This probabilistic picture of independent successive scatterings breaks
down at very high incident energies \cite{LPM1,LPM2,LPM3}. The squared 
amplitude
for $n$ scatterings now no longer factorizes into $n$ interactions;
instead, there is destructive interference, which for a regular medium
(crystal) leads to a complete cancellation of all photon emission except
for the first and last of the $n$ photons. This Landau-Pomeranchuk-Migdal 
(LPM) effect greatly reduces the radiative energy loss. The medium
produced in nuclear collisions is certainly not a regular crystal, so that 
here the cancellation becomes only partial.

\medskip

Nevertheless, the rates become considerably modified. The incoherent radiative
energy loss
\be
-{dE \over dz} \simeq {3 \alpha_s \over \pi} {E \over \lambda},
\label{9f}
\ee
is through interference replaced by the coherent form
\be
-{dE \over dz} \simeq {3 \alpha_s \over \pi} \sqrt{{\mu^2 E \over \lambda}}.
\label{10}
\ee
The former grows linearly with the energy of the passing charge and decreases
as its mean free path increases, i.e., if there is less scattering. 
In the latter the screening mass $\mu$ specifies in addition how much of the 
medium affects the jet and hence determines the amount of destructive 
interference. The crucial quantity is thus the {\sl transport coefficient}
$\hat q = \mu^2/\lambda$; it specifies the overall energy loss of the jet,
and it is much higher in a hot quark-gluon plasma than in a cool medium
of confined constituents.

\medskip

\begin{figure}[htb]
\centerline{\psfig{file=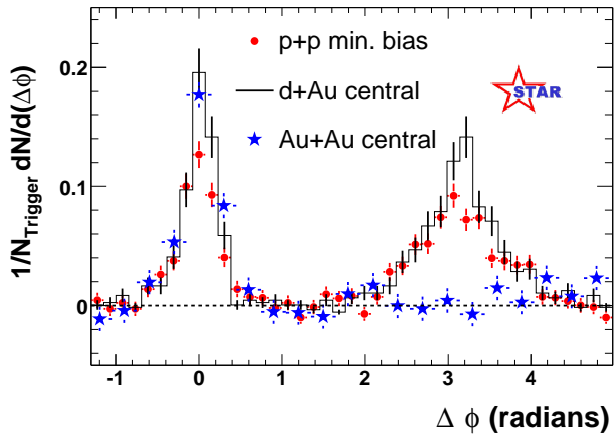,width=9cm}}
\caption{Azimuthal distribution of hadrons with transverse momenta above
5 GeV in $p-p$, $d-Au$ and $A-A$ collisions 
\cite{awayside1,awayside2,awayside3}.}
\label{azi}
\end{figure}

Estimates of the jet energy loss predicts for a QGP at $T = 250$ MeV 
some 3 GeV/fm, while cold nuclear matter only gives about 0.2 GeV/fm 
\cite{Schiff} -- more than an order of magnitude less. Such a difference 
in jet quenching has in fact been observed at the BNL Relativistic
Heavy Ion Collider RHIC.
In proton-proton collisions, a high transverse momentum particle is in
general accompanied by another such particle flying in the opposite
direction, to balance the overall transverse momentum (``back-to-back''
jets). In $d-Au$ collisions, the balancing jet has to traverse normal
nuclear matter to ``get out'', and this is found to have rather little
effect. In $Au-Au$ interactions, on the other hand, the balancing jet
has to pass through the produced hot QGP (if there is such a medium),
and this should lead to strong suppression. In Fig.\ \ref{azi} it is
seen that this is indeed observed -- the ``away-side'' jet has 
essentially become fully quenched.  

\medskip

The interpretation of this result in terms of jet quenching is further
supported by the observation that hard transverse photons are produced
in nucleus-nucleus collisions at the rate predicted from proton-proton
results scaled by the number of collisions; in contrast, hard hadron
production is reduced up to a factor five. The hadrons are quenched in
the QGP, while the photons are not affected by a strongly interacting
medium.

\medskip 

Numerous other features of jet production were studied at RHIC and
described in terms of QCD-based models; for more details, see e.g.\
\cite{Jai}. They all indicate that high energy nuclear collisions
indeed produce a very dense and very strongly interacting medium.
Hopefully the advent of further data from the LHC will eventually 
also lead to a quantitative comparison with 
{\sl ab initio} QCD calculations.

\section{\large Horizons}

The vast expanses of the universe, combined with its finite age, make it
not so surprising that there are regions even today which can never have
exchanged information with each other, which are causally not connected. 
Nevertheless the
cosmic microwave background radiation observed from these regions is of
the same universal temperature of some 2.7$^{\circ}$ Kelvin, up to better
than one part in tenthousand. How could this arise, how can regions be in 
appearant equilibrium with each other, even though they cannot communicate?
That is what cosmologists call their {\sl horizon problem}, and they conclude
that in classical Big Bang theory, there is no physical process which can
make the temperature so uniform. 

\medskip

We want to close our analysis of strongly interacting matter by noting that
such a horizon problem also occurs in the study of high energy collisions.
It made its first appearance when it was found that the relative abundances
of hadrons were essentially
the same in $e^+e^-$ annihilation as in high energy nuclear 
collisions. In the latter, one can imagine some kind of thermalisation 
through multiple parton interactions, but in $e^+e^-$ annihilation both
the number of partons and that of hadrons is so small that any kinetic
equilibration is ruled out. Hagedorn concluded that the hadrons are simply
``born in equilibrium''.

\medskip

\begin{figure}[h]
\centerline{\psfig{file=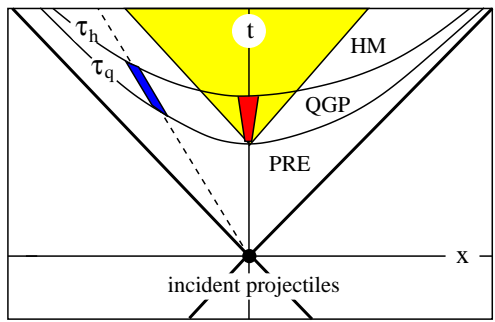,width=6.5cm}}
\caption{The evolution of a high energy nucleus-nucleus collision}
\label{evo}
\end{figure}

The canonical view of the evolution of a high energy nuclear collision is
illustrated in Fig.\ \ref{evo}. The projectiles pass through each other,
leaving behind an excited vacuum, which after a brief equilibration time
$\tau_q$ forms a quark-gluon plasma. This subsequently, after a further
time $\tau_h$, hadronises to form the observed secondaries. The evolution
is generally assumed to be boost-invariant; this means that in each local
rest frame, the same times $\tau_q$ and $\tau_h$ govern the development. 
As a result, the QGP bubble at mid-rapidity ($y=0$) and that at some
larger rapidity are not in causal contact; if the corresponding hadronisation
times are the same, it is not because the two systems ``thermalised'',
established a mutual equilibrium. They don't even know of each other's 
existence. 

\medskip

There seems to be only one way out of this dilemma: The hadron production
is due to a local stochastic process. In other words, at each space-time
point, hadronisation occurs through randomly chosing one out of the entire
set of allowed states -- allowed by the boundary conditions of the problem.
The set of all the hadrons obtained in this way looks ``thermal'', and
it is indeed the same set as would have been obtained through kinematic
thermalisation starting from a given non-equilibrium configuration. This
implies a form of stochastic equivalence principle. Just as one cannot
tell, by Einstein equivalence, if one is in a gravitational field or in
a rocket experiencing the same acceleration, so one cannot tell if a
given ensemble of states was obtained through kinematic equilibration
or through throwing dice. Once it's ``thermal'', it no longer remembers
how it got there.

\medskip

A model of this type was proposed five years ago \cite{CKS}, based on
the stochastic radiation seen by an accelerating observer \cite{Unruh}. 
The ``Unruh'' temperature obtained in this way is specified by the
acceleration, which in turn is determined by the string tension. And
such a scheme does in fact reproduce the universal hadrosynthesis
pattern observed in high energy collisions \cite{BCMS}. Further
work along these lines is in progress. 

\medskip

All in all, we can thus conclude that the little bang of nuclear
collisions perhaps has more features in common with its big brother
than we had bargained for. But it does leave us with many interesting
problems still to be solved. For more details, see Ref.\ \cite{HS},
where also references are given in a more systematic fashion; here
I have concentrated on some indicative works only.

\vskip1cm

\centerline{\large \bf Acknowledgement}

\bigskip

It is a great pleasure for me to thank Xin-Nian Wang and Nu Xu 
(both of LBL Berkeley) for making the {\sl 5th Berkeley School on 
Collective Dynamics in High Energy Collisions} a most stimulating,
informative and educational meeting.

\end{document}